\documentclass[letterpaper, 10 pt, conference]{ieeeconf}  
\usepackage{graphicx} 
\usepackage{caption}
\usepackage{subcaption}
\graphicspath{{./EDA/}}
\usepackage{multirow}
\usepackage{booktabs}
\usepackage{booktabs}
\usepackage{tabularx}
\usepackage{lipsum}
\usepackage{flushend}
\usepackage{float}
\usepackage{makecell, tabularx}
\newcolumntype{L}{>{\raggedright\arraybackslash}X}
\usepackage{siunitx}
\usepackage{booktabs,ragged2e}

\usepackage[flushleft]{threeparttable}

\IEEEoverridecommandlockouts              
\usepackage[utf8]{inputenc}
\usepackage[T1]{fontenc}

\title{\bf
Predicting Development of Chronic Obstructive Pulmonary Disease \\and its Risk Factor Analysis
}

\author{Soojin Lee$^{1, 2}$, Ingu Sean Lee$^{1}$, and Samuel Kim$^{1}$ 
\thanks{$^{1}$ Cipherome Inc. San Jose, U.S.A. {\tt\small \{rachell.lee, sean.lee, sam.kim\}@{cipherome.com}}}%
\thanks{$^{2}$Sejong University, Seoul, South Korea}%
}

\begin{document}

\maketitle
\thispagestyle{empty}
\pagestyle{empty}

\begin{abstract}
Chronic Obstructive Pulmonary Disease (COPD) is an irreversible airway obstruction with a high societal burden. Although smoking is known to be the biggest risk factor, additional components need to be considered. In this study, we aim to identify COPD risk factors by applying machine learning models that integrate sociodemographic, clinical, and genetic data to predict COPD development. 
\\ \\
\indent \textit{Clinical relevance}— This study assessed the risk factors of COPD in sociodemographic, clinical, and genetic data. We have determined that sociodemographic factors are highly associate to the development of COPD. 
\end{abstract}

\section{INTRODUCTION}
Chronic Obstructive Pulmonary Disease (COPD) affects more than 15 million Americans, with over 150,000 deaths annually, making it the sixth leading cause of death. Despite the high mortality and multi-factorial nature of COPD, few studies evaluate the risk factors, other than smoking, associated with COPD.  Studies have examined the patterns observed in COPD patients and have identified the patterns of multi-morbidity and polypharmacy in COPD~\cite{hanlonExaminingPatternsMultimorbidity2018}, while others have shown that sociodemographic factors~\cite{wang2018prevalence} and genetic variants~\cite{sakornsakolpat2019genetic} contribute to the development of COPD. The China Pulmonary Health (CHP) study was a large cross-sectional and multi-center study with subjects from ten different regions of China that  assessed the prevalence and risk factors of COPD in China~\cite{wang2018prevalence}. They analyzed the prevalence of risk factors in individuals with and without COPD, revealing that smoking, underweight, parental history of respiratory disease, and low education were major risk factors for COPD. 

To assess genetic factors, Sakornsakolpat et al~\cite{sakornsakolpat2019genetic} performed genome-wide association study (GWAS) to identify loci associated with COPD or lung function. Similarly, recent studies have performed GWAS to identify genetic risk loci~\cite{cho2014risk} associated with COPD. Analyses compared control and COPD groups combining data from cohorts such as COPDGene, ECLIPSE, NETT/NAS, and Norway GenKOLS studies. In addition to alpha-1 antitrypsin (A1AT), one of the first genes identified to be associated with COPD, novel genetic variants were found. These studies have shown insight into a broad array of risk factors attributed to the development of COPD. 

Previous studies are limited in that they focused either on sociodemographic or genetic factors, not both, when assessing the risk of COPD. COPD is a complex, multi-factoral disease, and all relevant risk factors should be considered simultaneously. In addition, most studies are cross-sectional and do not analyze the development or progression of the disease, which is crucial for chronic diseases. 

In this paper, we aim to determine risk factors for COPD development by evaluating comprehensive medical data including sociodemographic, clinical and genetic data. By longitudinal observation of these data, we hope to determine which factors correlate most to COPD development. We will use our proprietary research analysis platform, COMPASS, which analyzes medical data regardless of their source, to extract data and analyze important features. By identifying modifiable and non-modifiable risk factors, our goal is to enable early detection of COPD. This will lead to prompt management and ultimately COPD prevention. 

\section{Data}
\subsection{Data Source}

We used data from the UK Biobank~\cite{allen2012uk}, a large-scale biomedical database containing in-depth genetic, clinical, and sociodemographic details from 502,527 voluntary participants. We selected sociodemographic and clinical factors for COPD risk factors from Hanlon et al's study~\cite{hanlonExaminingPatternsMultimorbidity2018}. For the genomic factors, we used genomic features known to be associated with lung function and COPD. From the list of genes shown in the review series~\cite{RobertHall2018}, we chose 10 single nucleotide polymorphism (SNP) from 7 genes using an array-based data.

According to Hanlon et al, material deprivation is one of the sociodemographic risk factors for COPD development. Townsend scores is a measure of this and is calculated using a combination of four census variables for a geographical area. The scores are obtained from participant postcodes to provide an area-based measure of socioeconomic deprivation. A greater Townsend index score implies a greater degree of deprivation. Smoking status and frequency of alcohol intake are also known risk factors. Physical Activity was self-reported and classified into 6 groups. Additional physical measures such as height and weight were collected. In the UK Biobank, all disease, health condition and medications at the time of the assessment center visit, were self-reported.\par

UK Biobank provides a record-level hospital inpatient data. We used the International Classification of Disease codes (ICD-9 and ICD-10) to acquire the exact date of COPD diagnosis. \par

\subsection{Problem definition}

Participants diagnosed with COPD, chronic bronchitis or emphysema were classified as 'self-reported COPD' and others were classified as 'no-COPD' (n=56,231). To investigate the development of COPD, participants from the 'no-COPD' group were investigated using the date of diagnosis extracted from ICD codes. If the participants' diagnosis date of COPD was later than that of assessment center visit, where the initial report of COPD was executed, they were categorized as 'no-COPD to COPD' (n=1,961). Participants who did not have diagnosis date or had an earlier diagnosis date compared to center visit remained as 'no-COPD' (n=54,252).

Cardiovascular conditions were categorized into 7 conditions. If a participant self-reported any of the conditions among hypertension, coronary heart disease, diabetes, stroke/TIA, atrial fibrillation, heart failure, and/or peripheral vascular disease, they were labeled as having cardiovascular conditions. \par
Medication data was grouped into 4 drug classes; oral steroids, selective serotonin reuptake inhibitors (SSRIs), non-steroidal anti-inflammatory drugs (NSAIDs), and anti-platelet agents. Based on the reported medication, participants were labeled yes/no in each class of drugs.  

\section{Experiments}
\subsection{Cohort Selection}
\begin{figure}[b!]
  \centering
      \includegraphics[width=8.6cm]{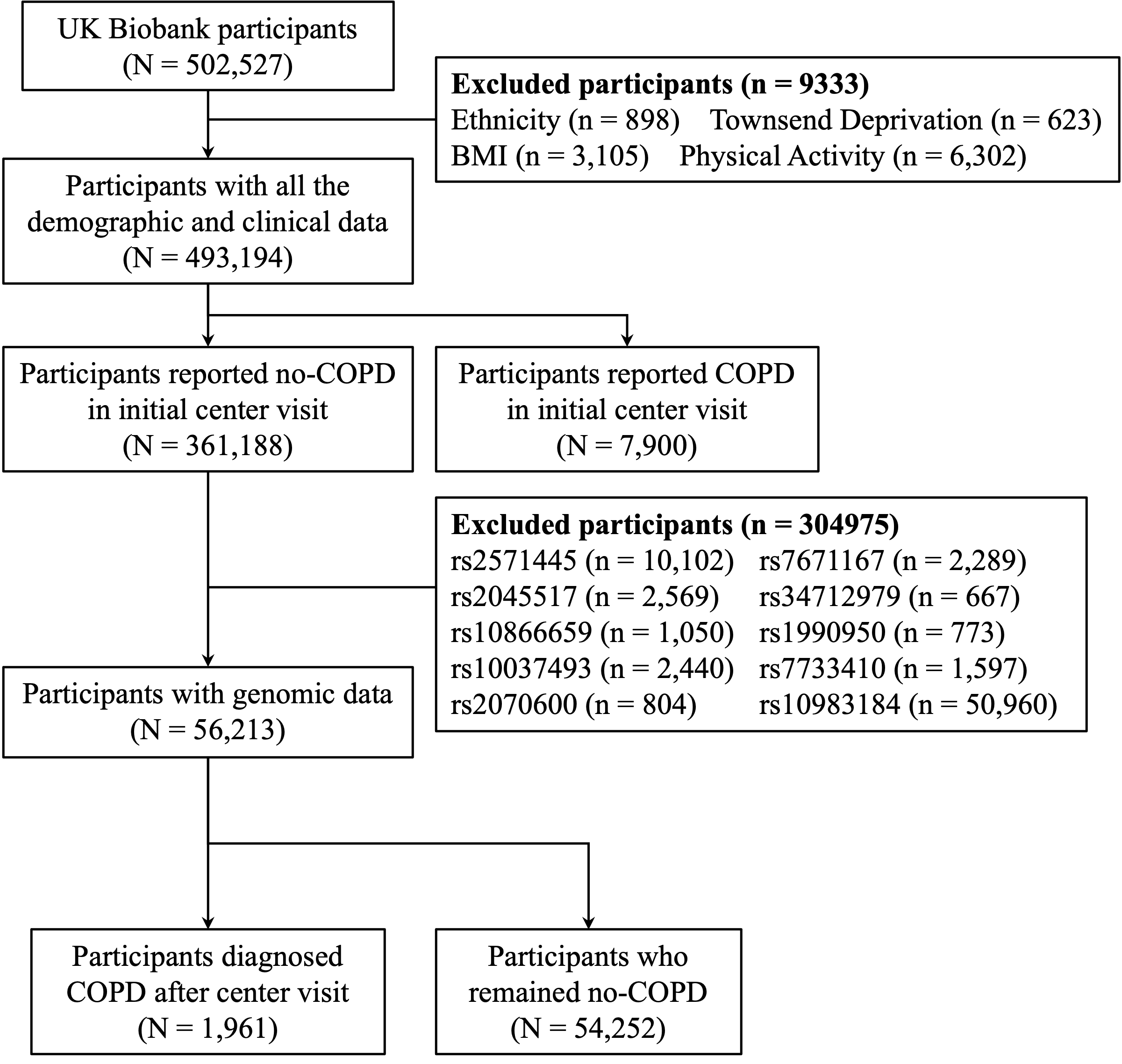}
  \caption{Flow diagram of participants}
  \label{figure:diagram}
\end{figure}

Fig.~\ref{figure:diagram} is a flow diagram of cohorts. At initial center visit, 502,527 participants enrolled. Participants without information on ethnicity, BMI, Townsend deprivation score, physical activity were excluded (n=9,333). While 7,900 out of 493,194 participants self-reported COPD, 361,188 reported no-COPD. To assess which risk factors correlated with the development of COPD, participants who reported no-COPD at initial center visit were the focus of this study. From no-COPD participants (N=361,188), participants without genomic data were excluded. Among the participants with all genomic, sociodemographic and clinical data (N=56,213), 54,252 participants remained without COPD and 1961 participants developed COPD. The latter group was the focus of our study.

\subsection{Exploratory Analysis}

\begin{figure*}[ht]
\centering
  \begin{subfigure}[]{0.3\linewidth}
    \includegraphics[width=\linewidth]{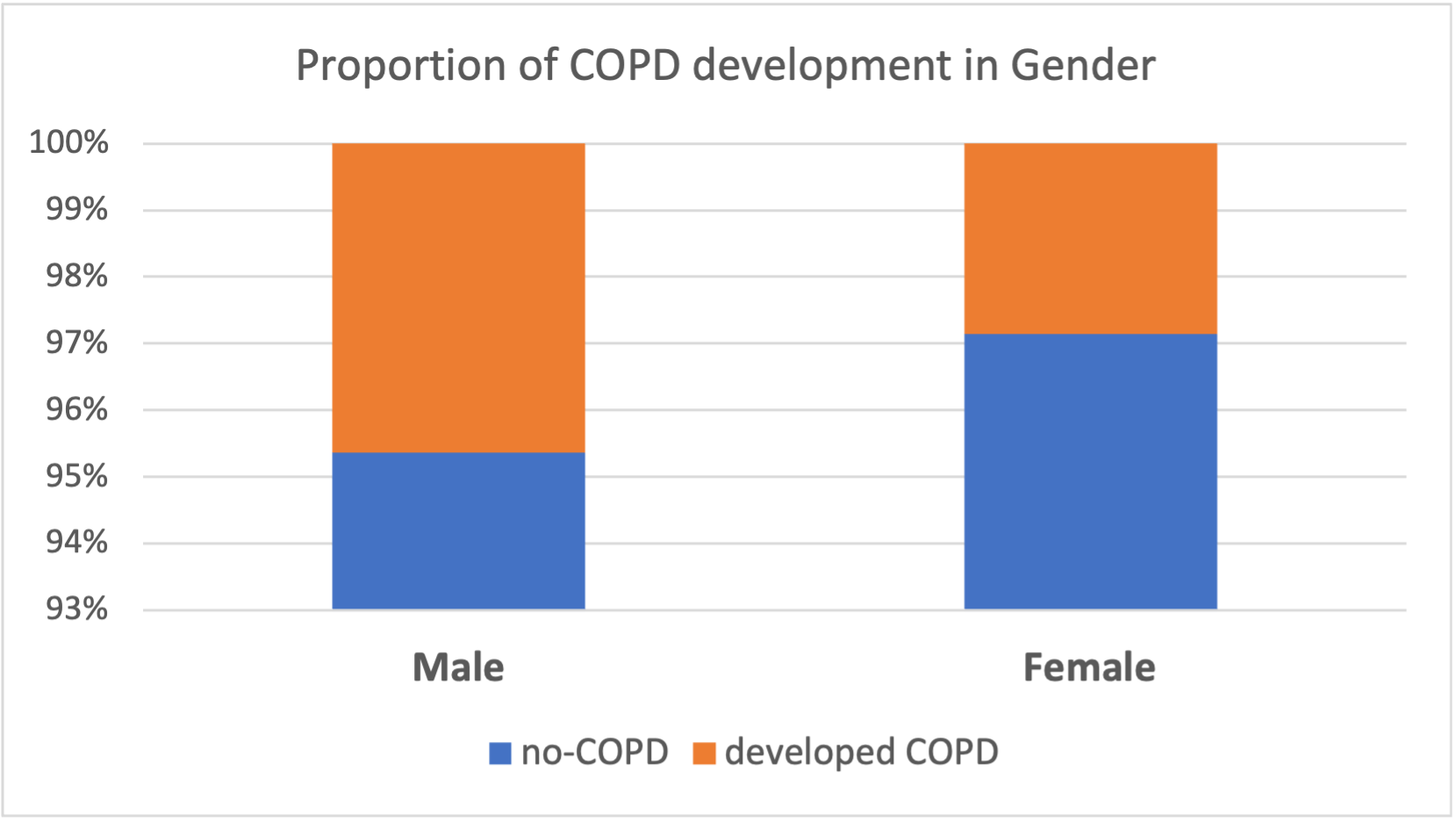}
     \caption{Gender}
     \label{fig:gender}
  \end{subfigure}
  \begin{subfigure}[]{0.3\linewidth}
    \includegraphics[width=\linewidth]{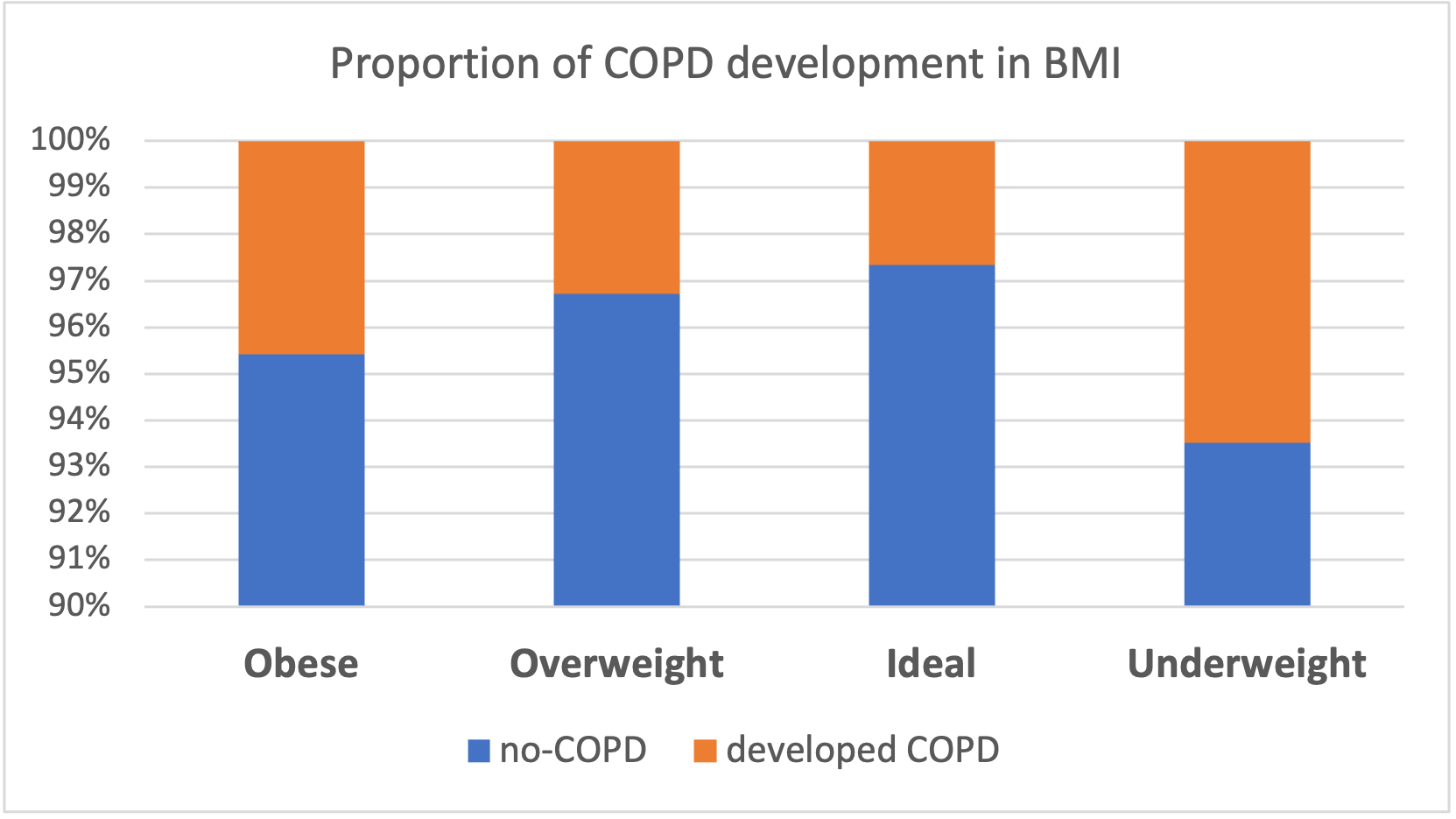}
     \caption{BMI}
  \end{subfigure}
  \begin{subfigure}[]{0.3\linewidth}
    \includegraphics[width=\linewidth]{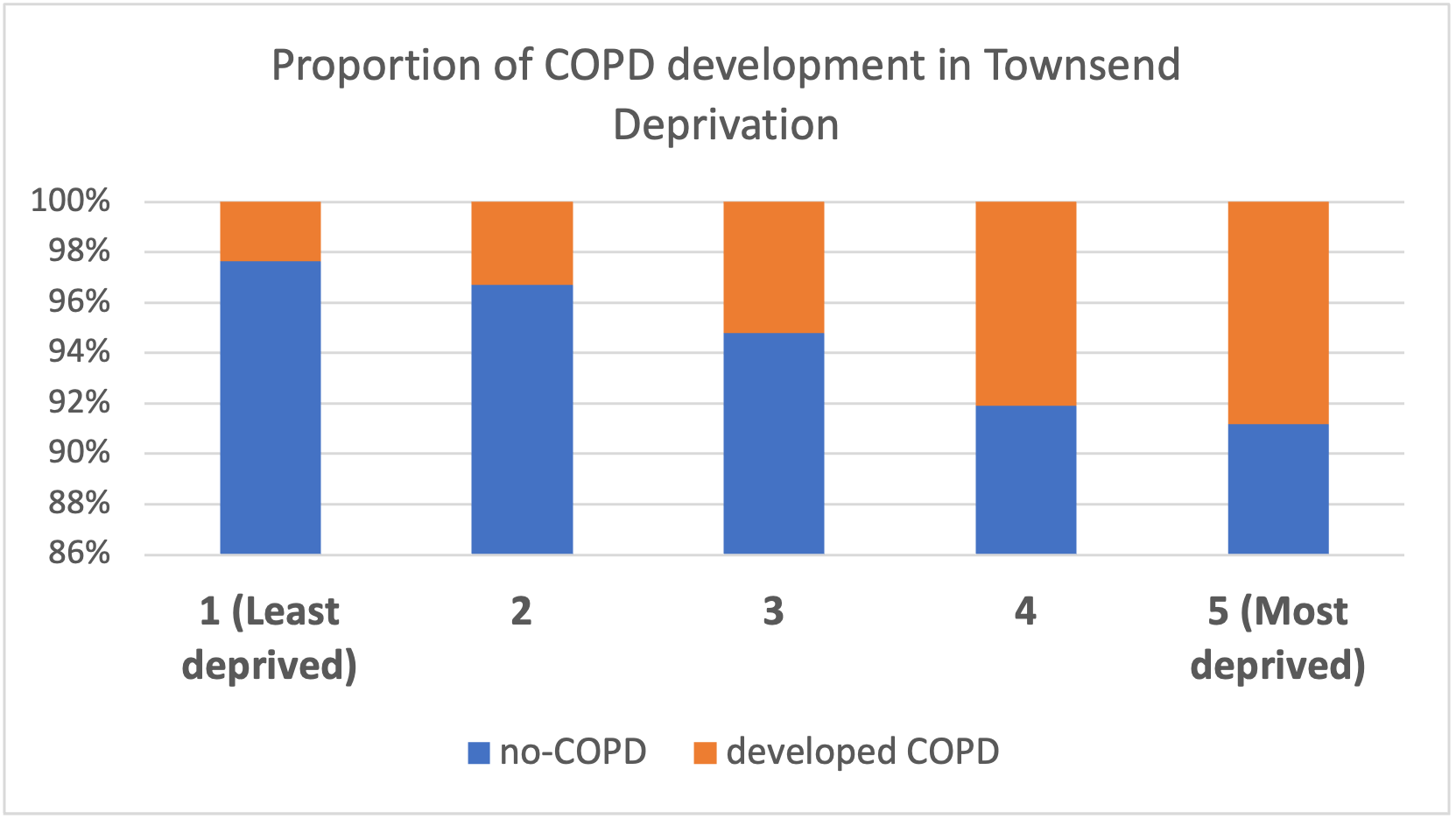}
    \caption{Socioeconomic Deprivation}
  \end{subfigure}
  \begin{subfigure}[]{0.3\linewidth}
    \includegraphics[width=\linewidth]{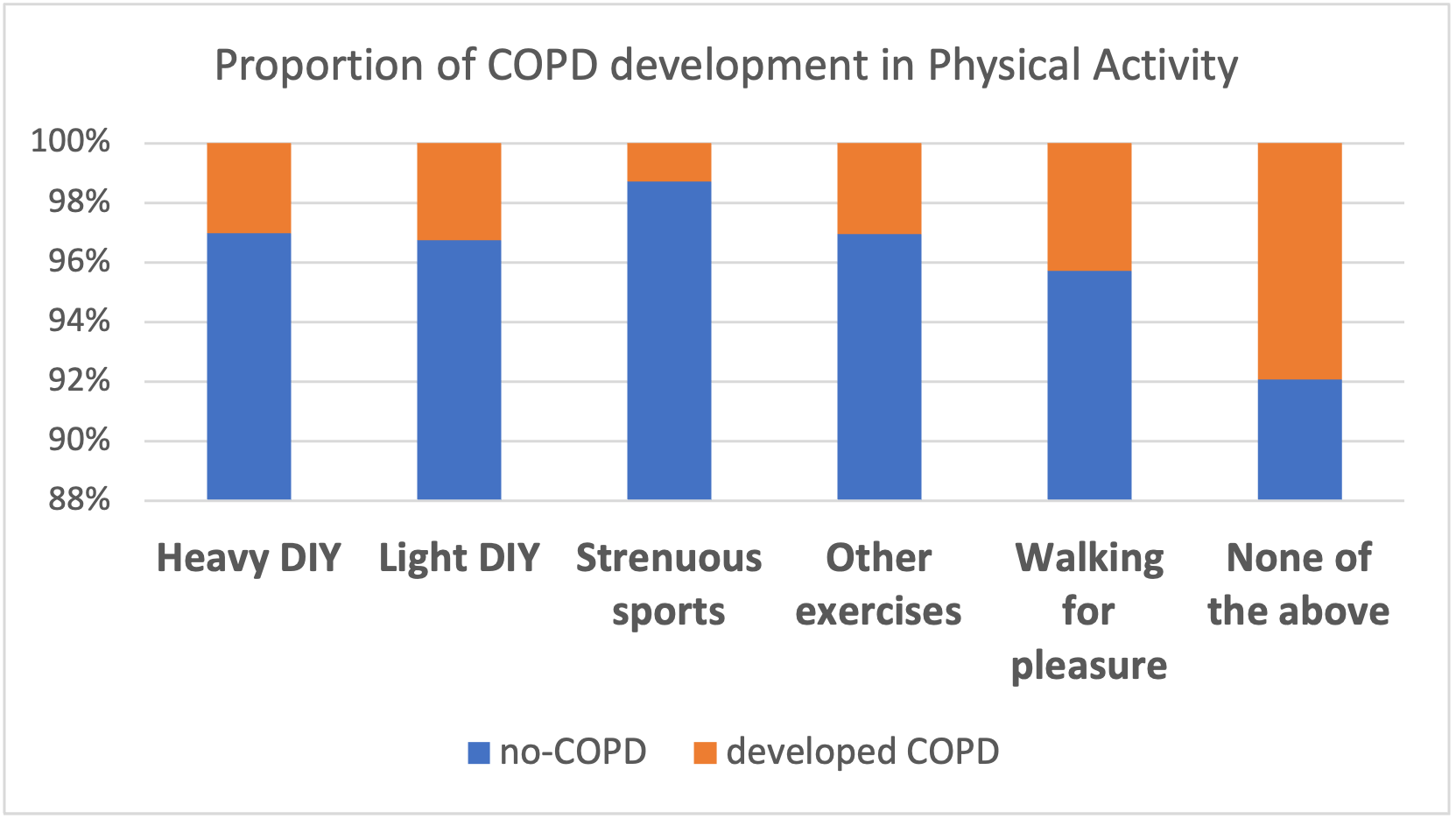}
    \caption{Physical Activity}
  \end{subfigure}
  \begin{subfigure}[]{0.3\linewidth}
    \includegraphics[width=\linewidth]{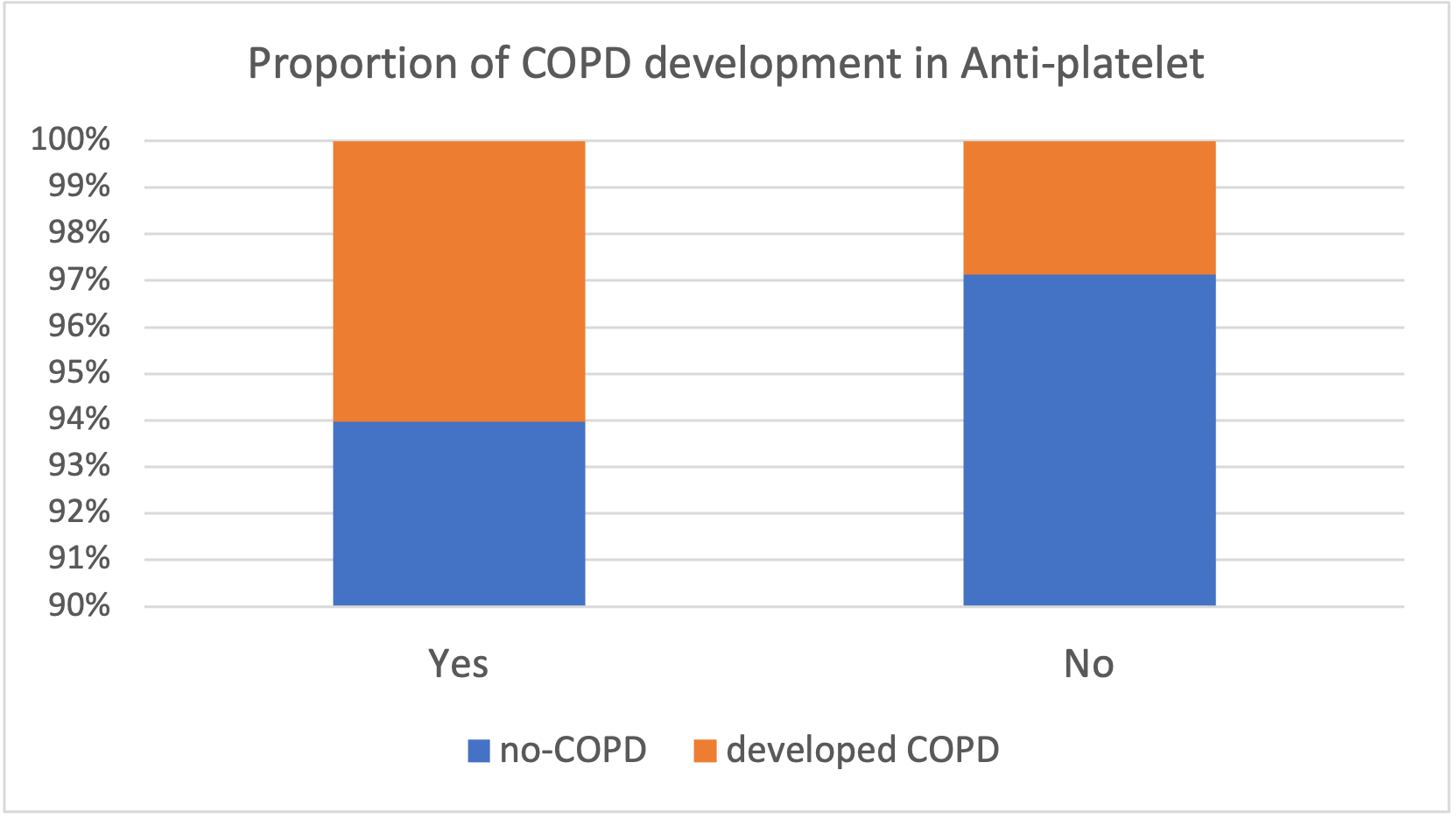}
    \caption{Anti-platelet}
  \end{subfigure}
  \begin{subfigure}[]{0.3\linewidth}
    \includegraphics[width=\linewidth]{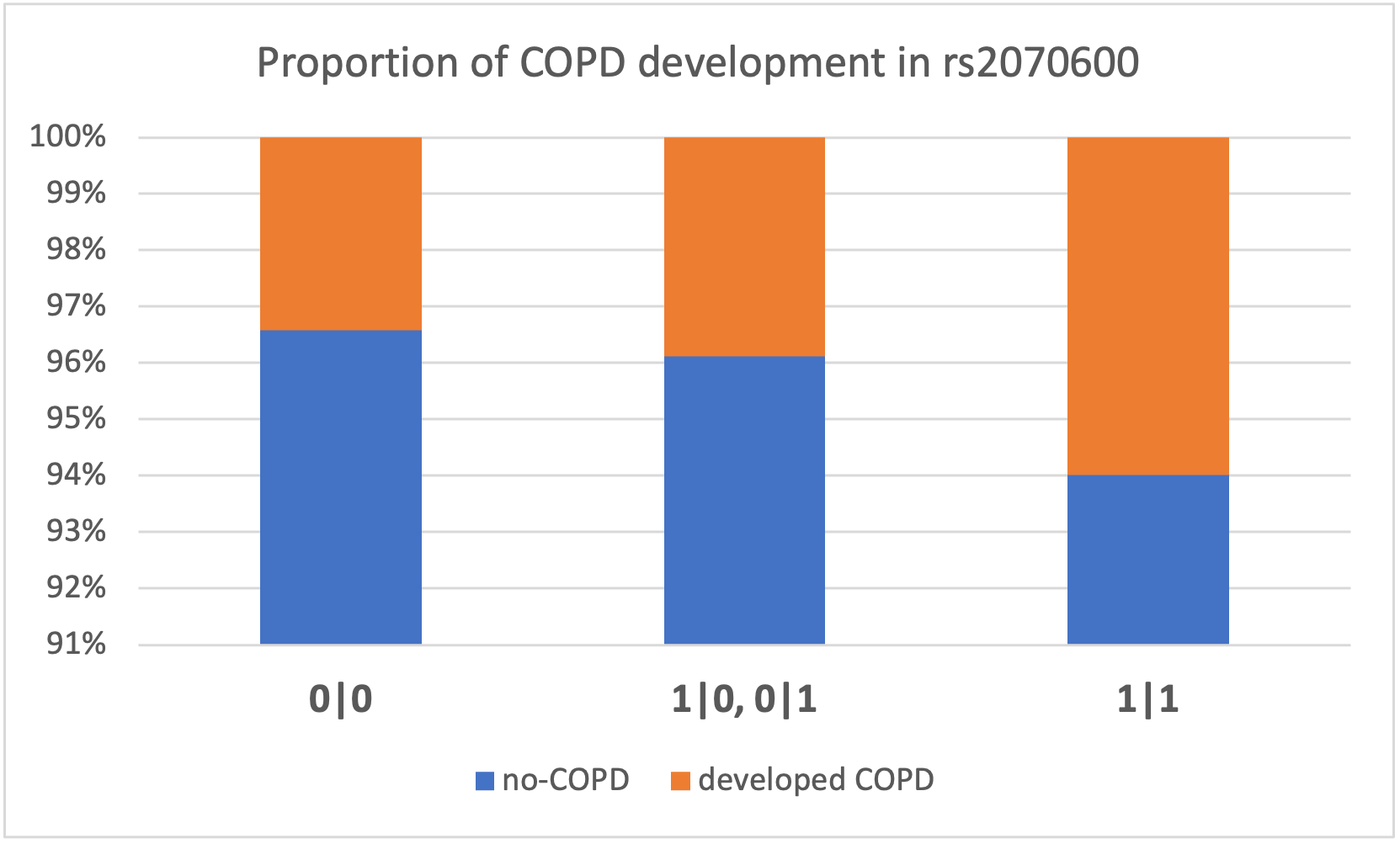}
    \caption{rs2070600}
  \end{subfigure}
   \begin{subfigure}[]{0.4\linewidth}
    \includegraphics[width=\linewidth]{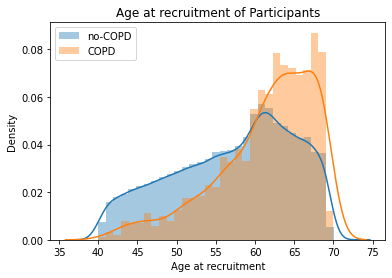}
    \caption{Age}
  \end{subfigure}
  \begin{subfigure}[]{0.4\linewidth}
    \includegraphics[width=\linewidth]{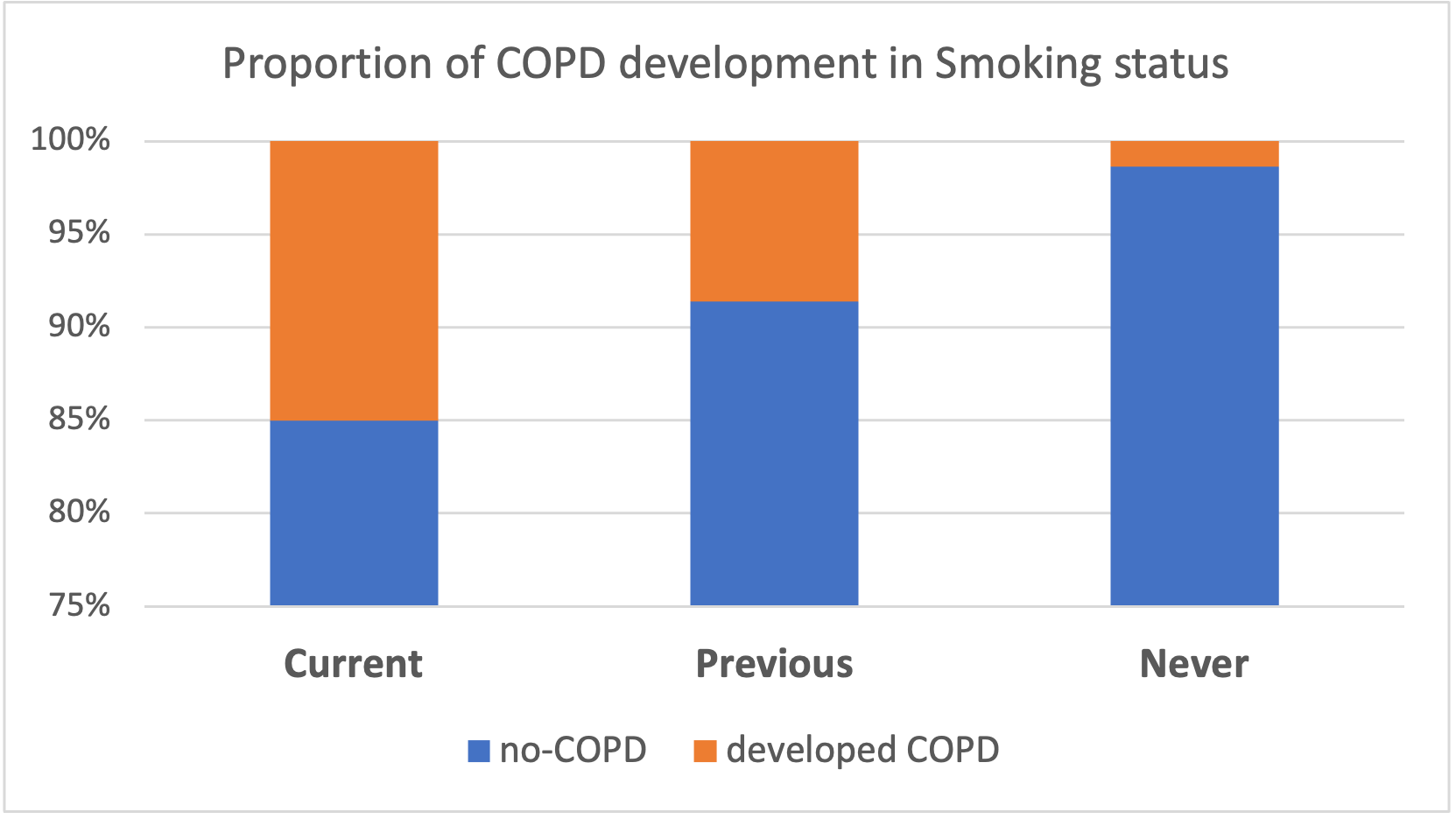}
    \caption{Smoking}
  \end{subfigure}
  \caption{Compare proportion of COPD development in each factor}
  \label{fig:eda}
\end{figure*}

Fig.~\ref{fig:eda} provides exploratory analysis results. To identify patterns of 'no-COPD to COPD' group, we used stacked bar plot and histogram. Proportion of 'no-COPD to no-COPD' and 'no-COPD to COPD' in each factor are shown in the stacked bar chart. Factors that are represented in continuous values such as age are displayed by the distribution of each group.

To identify the patterns of participants who developed COPD, we compared the characteristics of those who developed COPD and who did not. The results indicate that participants who developed COPD tended to be male, older, underweight, more materially deprived and less physically active. In addition, a higher proportion of those who developed COPD were taking anti-platelet. History of smoking was most prevalent in participants who developed COPD. From list of genetic variants, the rs2070600 variant was associated with COPD development.

\subsection{Experimental setup}

With a total of 23 features, we trained XGBoost (XGB), Logistic Regression (LR), Naive Bayes (NB), and Random Forest (RF) to predict the development of COPD. Table ~\ref{table:feature} shows the list of features and categorization of features. 

    \begin{table}[ht]
    \setcellgapes{2pt}
    \makegapedcells
\centering
\caption{Feature categorization}
\begin{tabularx}{\linewidth}{ l | L  }
\hline \hline
    \textbf {Sociodemographic (A)}    &  Gender, Recruitment Age, Ethnicity, Townsend Deprivation, Body Mass Index (BMI), Physical Activity, Smoking Status, Alcohol Frequency   \\ \hline
    \textbf {Clinical (B)}    &  Cardiovascular Conditions, Anti-platelet, Oral Steroids, SSRIs, NSAIDs   \\ \hline
    \textbf{Genetic (C)}     &  rs2571445, rs7671167, rs2045517,rs34712979, rs10866659, rs1990950, rs10037493,rs7733410, rs2070600, rs10983184   \\ \hline \hline
\end{tabularx}
\label{table:feature}
    \end{table}

We evaluated each model's performance using AUC-ROC curve and the value of AUC. After training the models with all the features, we conducted ablation studies to validate the effectiveness of feature categories with multiple settings 1) only with sociodemographic 2) only with clinical 3) only with genetic variants 4) combination of sociodemographic and clinical 5) combination of sociodemographic and genetic variants 6) combination of clinical and genetic variants 7) combination of sociodemographic, clinical, and genetic variants. Furthermore, we used SHapley Additive exPlanations (SHAP)~\cite{lundberg2017unified}, an interpretable machine learning model to compute the contributions of each feature to the prediction on the development of COPD.

\subsection{Results}

\begin{table}[ht]
\begin{threeparttable}
\caption{Comparison of AUC score in different settings}
\label{tab:2}
\setlength\tabcolsep{2pt} 

\begin{tabular*}{\columnwidth}{@{\extracolsep{\fill}} l cccc @{}}
\toprule
     Feature Setting & \multicolumn{4}{c}{Model AUC} \\ 
\cmidrule{2-5}
    & XGB & LR & NB & RF\\
\midrule
     A  & 0.815 & 0.814 & 0.791 & 0.713 \\
     B  & 0.632 & 0.63 & 0.622 & 0.626 \\
     C  & 0.519 & 0.497 & 0.51 & 0.507 \\
\addlinespace
     A+B  & 0.818 & 0.818 & 0.783 & 0.755\\
     A+C  & 0.813 & 0.814 & 0.792 & 0.771 \\
     B+C   & 0.633 & 0.626 & 0.621 & 0.554 \\
\addlinespace
     A+B+C  & 0.817 & 0.818 & 0.783 & 0.787 \\
\bottomrule
\label{table:auc}
\end{tabular*}
\begin{tablenotes}
      \small
      \item Note: A, B, C can be found from Table \ref{table:feature}. '+' is to show the combination of categories.
\end{tablenotes}
\end{threeparttable}

\end{table}
    
Table~\ref{table:auc} shows the AUC score of each machine learning models in 7 different settings. Out of 3 settings, sociodemographic, clinical, and genetic, sociodemographic scored the highest in AUC. Clinical factors showed a relatively stable performance in all machine learning models. However, having genetic factors as predictor variables showed low performance in overall models. In addition, combinations that include sociodemographic features tend to assure an AUC score of 0.813. Reversely, adding genetic factors to clinical factors seem to have negative effect on the prediction. The two settings that performed the best AUC of 0.818 was combining all the features of sociodemographic, clinical, and genetic factors or combining sociodemographic and clinical factors. Among 4 machine learning models, XGB performed the best in most of the settings. While XGB and LR scored the same AUC of 0.818 with the use of sociodemographic and clinical factors, LR scored the highest in sociodemographic, clinical, and genetic setting.

On each setting, we used SHAP to all machine learning models to determine what the important features are and how each feature affects the prediction results. We used barplot and beeswarm plots to summarize the entire distribution of SHAP values for each feature. 

\begin{figure}[ht]
  \centering
      \includegraphics[width=8cm]{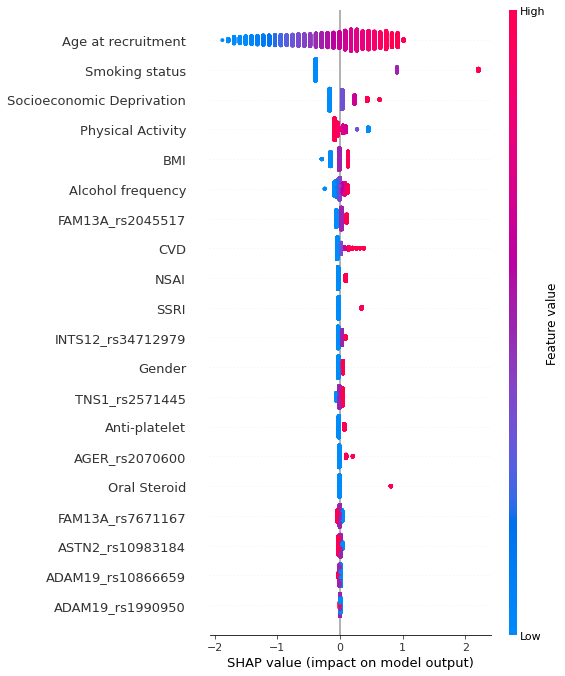}
  \caption{SHAP value plot in Sociodemographic, Clinical, Genetic setting}
  \label{figure:shap}
\end{figure}

As shown in Fig.~\ref{figure:shap}, among all the features, sociodemographic features contributed the most to COPD prediction. The features were shown in order of importance from top to bottom, indicating age at recruitment, smoking status, and socioeconomic deprivation had the highest impact. Other sociodemographic features that showed significance were physical activity and BMI. In particular, for smoking status the higher the value, which meant the participant had a history of smoking, the higher the development of COPD. Also, the older the participant, the higher the vulnerability to COPD. In addition, patients living in more deprived areas tended to develop COPD. Participants who were less physically active and had high BMI were also likely to develop COPD. The next feature category that showed impact on model prediction was clinical factors. The existence of cardiovascular conditions had impact on the models' output. Also, intake of NSAIDs or SSRIs played a part in predicting the development of COPD. From the SHAP values of genetic features, we could not assume that genetic variants are associated with the development of COPD. Although showing less effect, rs2045517 and rs34712979 were associated with COPD development.

\subsection{Limitations}
 UK Biobank participants tended to be healthier with a lower frequency of COPD development. Demographic and clinical data relied on participants' self-report and thus may have been susceptible to bias or inaccuracy. In addition, because the genes were randomly selected, there could have been other genomic variants that had a greater impact on COPD. In addition, people who have gene variants associated with COPD tended to be diagnosed at an early age. Due to this propensity, genomic variants may not have shown influence in the development of COPD.

\section{Conclusion}
Participants who developed COPD were predominantly male, older, underweight or obese, economically deprived, less physically active, on anti-platelet agents, and expressed the rs2070600 variant. With domain knowledge on COPD risk factors, we applied explanatory machine learning method to verify major risk factors of COPD. As expected, we found that age, smoking status and socioeconomic deprivation were associated with COPD development. In contrast, genetic factors identified from previous studies did not show strong association in COPD development. Compared to genetic factors, medication history and medical conditions were more reliable indicators of COPD development. Having comprehensively examined of all the aforementioned risk factors, we concluded that lifestyle factors significantly impact COPD development. In other words, the risk of having COPD can be reduced by improving lifestyle. With the use of machine learning technology, further research should examine the relationship between medical conditions or drug intake and COPD to provide accurate diagnosis and effective treatment.


\bibliographystyle{IEEEbib}
{\bibliography{mybib}}

\begin{thebibliography}{1}

\bibitem{hanlonExaminingPatternsMultimorbidity2018}
Peter Hanlon, Barbara~I Nicholl, Bhautesh~Dinesh Jani, Ross McQueenie, Duncan
  Lee, Katie~I Gallacher, and Frances~S Mair,
\newblock ``Examining patterns of multimorbidity, polypharmacy and risk of
  adverse drug reactions in chronic obstructive pulmonary disease: a
  cross-sectional {UK} {Biobank} study,''
\newblock {\em BMJ Open}, vol. 8, no. 1, pp. e018404, Jan. 2018.

\bibitem{wang2018prevalence}
Chen Wang, Jianying Xu, Lan Yang, Yongjian Xu, Xiangyan Zhang, Chunxue Bai,
  Jian Kang, Pixin Ran, Huahao Shen, Fuqiang Wen, et~al.,
\newblock ``Prevalence and risk factors of chronic obstructive pulmonary
  disease in {C}hina (the {C}hina {P}ulmonary {H}ealth study): a national
  cross-sectional study,''
\newblock {\em The Lancet}, vol. 391, no. 10131, pp. 1706--1717, 2018.

\bibitem{sakornsakolpat2019genetic}
Phuwanat Sakornsakolpat, Dmitry Prokopenko, Maxime Lamontagne, Nicola~F Reeve,
  Anna~L Guyatt, Victoria~E Jackson, Nick Shrine, Dandi Qiao, Traci~M Bartz,
  Deog~Kyeom Kim, et~al.,
\newblock ``Genetic landscape of chronic obstructive pulmonary disease
  identifies heterogeneous cell-type and phenotype associations,''
\newblock {\em Nature genetics}, vol. 51, no. 3, pp. 494--505, 2019.

\bibitem{cho2014risk}
Michael~H Cho, Merry-Lynn~N McDonald, Xiaobo Zhou, Manuel Mattheisen, Peter~J
  Castaldi, Craig~P Hersh, Dawn~L DeMeo, Jody~S Sylvia, John Ziniti, Nan~M
  Laird, et~al.,
\newblock ``Risk loci for chronic obstructive pulmonary disease: a genome-wide
  association study and meta-analysis,''
\newblock {\em The lancet Respiratory medicine}, vol. 2, no. 3, pp. 214--225,
  2014.

\bibitem{allen2012uk}
Naomi Allen, Cathie Sudlow, Paul Downey, Tim Peakman, John Danesh, Paul
  Elliott, John Gallacher, Jane Green, Paul Matthews, Jill Pell, et~al.,
\newblock ``Uk biobank: Current status and what it means for epidemiology,''
\newblock {\em Health Policy and Technology}, vol. 1, no. 3, pp. 123--126,
  2012.

\bibitem{RobertHall2018}
Robert Hall, Ian Hall, and Ian Sayers,
\newblock ``Genetic risk factors for the development of pulmonary disease
  identified by genome‐wide association,''
\newblock {\em Respirology}, vol. 24, 11 2018.

\bibitem{lundberg2017unified}
Scott~M Lundberg and Su-In Lee,
\newblock ``A unified approach to interpreting model predictions,''
\newblock {\em Advances in neural information processing systems}, vol. 30,
  2017.

\end{thebibliography}

\section*{ACKNOWLEDGMENT}
This research has been conducted using the UK Biobank Resource under Application Number 52031.

\end{document}